\documentclass[]{spie}  

\usepackage{amsmath,amsfonts,amssymb}
\usepackage{graphicx}
\usepackage[colorlinks=true, allcolors=blue]{hyperref}

\title{The simulation framework of the timing-based localization for future all-sky gamma-ray observations with a fleet of Cubesats}

\author[a,b,c]{Masanori Ohno}
\author[d]{Norbert Werner}
\author[e]{Andr{\' a}s P{\' a}l}
\author[e]{L{\'a}szl{\'o} M{\'e}sz{\'a}ros}
\author[f]{Yuto Ichinohe}
\author[d,a]{Jakub {\v R}{\' i}pa}
\author[d]{Martin Topinka}
\author[d]{Filip Munz}
\author[a,g]{Gab{\' o}r Galg{\' o}czi}
\author[c]{Yasushi Fukazawa}
\author[c]{Tsunefumi Mizuno}
\author[c]{Hiromitsu Takahashi}
\author[c]{Nagomi Uchida}
\author[c]{Kento Torigoe}
\author[c]{Naoyoshi Hirade}
\author[c]{Kengo Hirose}
\author[c]{Hiroto Matake}
\author[h]{Kazuhiro Nakazawa}
\author[h]{Syohei Hisadomi}
\author[i]{Hirokazu Odaka}
\author[j]{Teruaki Enoto}
\author[k]{J\'an Hudec}
\author[k]{Jakub Kapus}
\author[l]{Martin Koleda}
\author[l]{Robert Laszlo}
\affil[a]{Institute of Physics, E\"otv\"os University, P\'azm\'any P\'eter s\'et\'any 1/A, Budapest, 1117, Hungary}
\affil[b]{MTA-E\"ot\"vos University Lend\"ulet Hot Universe Research Group, P\'azm\'any P\'eter s\'et\'any 1/A, Budapest, 1117, Hungary}
\affil[c]{School of Science, Hiroshima University, 1-3-1 Kagamiyama, Higashi-Hiroshima, Japan }
\affil[d]{Department of Theoretical Physics and Astrophysics, Faculty of Science, Masaryk University, Kotl\'a\v{r}sk\'a 2, Brno, 611 37, Czech Republic }
\affil[e]{Konkoly Observatory of the Hungarian Academy of Sciences, Konkoly-Thege ut 15-17, Budapest, 1121, Hungary}
\affil[f]{Department of Physics, Rikkyo University, Nishi Ikebukuro 3-34-1, Toshimaku, Tokyo 171-8501, Japan}
\affil[g]{Wigner Research Centre, Konkoly-Thege Miklós út 29-33, Budapest, 1121, Hungary}
\affil[h]{Department of Physics, Nagoya University, Furo-cho, Chikusa-ku, Nagoya, Aichi, Japan}
\affil[i]{Department of Physics, University of Tokyo, 7-3-1 Hongo, Bunkyo, Tokyo 113-0033, Japan}
\affil[j]{The Hakubi Center for Advanced Research and Department of Astronomy, Kyoto University, Kyoto 606-8302, Japan}
\affil[k]{Spacemanic s.r.o., Zámocká 5
811 01 Bratislava
Slovakia}
\affil[l]{Needronix s.r.o., Ilkovičova 3, 841 04 Bratislava, Slovakia}

\authorinfo{Further author information: (Send correspondence to Masanori Ohno)\\Masanori Ohno: E-mail: ohno@caesar.elte.hu}

\begin{document} 
\maketitle

\begin{abstract}
The timing-based localization, which utilize the triangulation principle with the different arrival time of gamma-ray photons, with a fleet of Cubesats is a unique and powerful solution for the future all-sky gamma-ray observation, 
 which is a key for identification of the electromagnetic counterpart of the gravitational wave sources. The Cubesats Applied for MEasuring and Localising Transients (CAMELOT) mission is now being promoted by the Hungarian and Japanese collaboration with a basic concept of the nine Cubesats constellations in low earth orbit. The simulation framework for estimation of the localization capability has been developed including orbital parameters, an algorithm to estimate the expected observed profile of gamma-ray photons, finding the peak of the cross-correlation function,  and a statistical method to find a best-fit position and its uncertainty. It is revealed that a degree-scale localization uncertainty can be achieved by the CAMELOT mission concept for bright short gamma-ray bursts, which could be covered by future large field of view ground-based telescopes. The new approach utilizing machine-learning approach is also investigated to make the procedure automated for the future large scale constellations. The trained neural network with 10$^6$ simulated light curves generated by the artificial short burst templates successfully predicts the time-delay of the real light curve and achieves a comparable performance to the cross-correlation algorithm with full automated procedures.

\end{abstract}


\section{INTRODUCTION}

The detection of the electromagnetic counterpart of the gravitational wave sources should be explored in the future multi-messenger astronomy since there is only one confirmation of the electromagnetic counterpart from GW 170817 despite that the improvement of the sensitivity of the gravitational wave telescope brought more than 50 gravitational wave detection during their 3rd observational run. The all-sky observations in gamma-ray energy band is one of the keys to identify the electromagnetic counterpart of the gravitational wave sources since gamma-ray bursts (GRBs) are one of the strong candidates of the gravitational wave source. 

The GRB is the most energetic explosion in the universe which is thought to be originated by the core-collapse of the massive star or merger of compact objects (e.g., neutron stars and black-holes) at the cosmological distance. These systems are thought to have an ultra-relativistic jet towards our line-of-sight and radiate strong gamma-rays in a short time scale, which is milliseconds to hundreds seconds. Especially, it is considered that the merger of the compact objects makes a specific population of the duration in a shorter regime ($<$ 2 seconds), which is referred to as the short-duration GRBs or simply short GRBs. The merger of the compact object is also a leading candidate of the gravitational wave sources. Therefore, it should be promising that the association between gravitational wave detection and the gamma-ray signal from short GRBs is discovered. 

The unique detection of the electromagnetic counterpart from GW 170817 accompanied weak gamma-ray signals and one of the interpretations is the off-axis jet model\cite{2017Natur.551...71T,2018ATel11619....1T}.
Association between strong gamma-rays from the on-axis jet of short GRBs and gravitational wave signals is really anticipated in the future observation run of the gravitational wave telescopes. Since there is no way to know when and where GRBs will happen in advance, the all-sky gamma-ray observations are essential to detect the gamma-ray signal from GRBs without any observational gaps due to earth occultation and operational constraints. Furthermore, in order to establish the association with the gravitational wave signal, which have a large localization uncertainty of more than tens square degrees and to perform more detailed follow-up observations by the ground-based optical telescope with large field-of-view with degree in the future, degree to sub-degree localization accuracy in gamma-ray observations is anticipated. 

The timing-based localization with a fleet of CubeSats is an unique and powerful solution to realize the all-sky observations at any time with a degree to sub-degree localization uncertainty.
The incoming photon direction can be constrained by the different arrival time of gamma-ray photons among multiple satellites in a different orbit by the simple triangulation principle\cite{2013ApJS..207...39H} ${\rm cos}\theta = c\delta t/D$, where c, $\delta t$, and D represent the speed of light, difference of the photon arrival time, and the distance of two satellites, respectively. Two satellites can constrain the photon direction in the annulus with cos$\theta$, and two annuli can be drawn by three satellites and two intersection areas are found to be the candidates of the photon direction. Finally, only one intersection area can be determined by adding another one 
non-coplanar satellite, by the Earth occultation technique or by the anisotropic response of a used detector.
The InterPlanetary Network (IPN) has utilized this principle to localize GRBs with satellites operated at a long distance. 

While, the concept of the timing-based localization with CubeSats utilize the satellites in low earth orbit. Increasing number of satellites and timing synchronization accuracy by the Global Positioning System (GPS) are expected to realize degree to sub-degree localization uncertainty. Cubesats Applied for MEasuring and LOcalising Transients (CAMELOT) is now being proposed as the Hungarian and Japanese collaboration CubeSats mission to realize the degree to sub-degree scale localization of the transients in the gamma-ray sky to identify the electromagnetic counterpart of the gravitational wave sources.
The mission concept of the CAMELOT is to fly nine-satellite constellations of 3-U CubeSats. Each satellite has four sets of CsI scintillators with the size of 150$\times$75$\times$5 mm$^3$ to maximize the effective area on the 3-U CubeSats platform. The scintillation light is readout by the Si-PM photon counter, which is named as the Multi-Pixel-Photon-Counter (MPPC) provided by the Hamamatsu photonics Japan. The detailed mission and detector concept are found in Werner et al. 2018\cite{2018SPIE10699E..2PW} and Ohno et al. 2018\cite{2018SPIE10699E..64O}, and two demonstration missions are now about to be launched. 

Our prototype detector including analog and digital electronics boards has been developed for the 1-U CubeSats platform. Two detectors will be launched as part of the payload module of Czech 3-U CubeSat mission, VZLUSAT-2. The flight model of our detectors have already been delivered to the satellite and passed various environmental tests such as the vibration and thermal vacuum tests and the satellite is now in the United States ready for launch in December. 

Another detector will be also launched as the payload module of 1-U CubeSat platform, which is developed by the collaboration with the Slovak company. This "GRBAlpha" mission is now under final environmental test phase, and it will be launched next year by Russian Soyuz rocket.
The mission and hardware concept will be verified by those 1-U demonstration missions. 

Another important study is the timing-based localization algorithm. The detailed concept of this algorithm was reported in Ohno et al. 2018. However, there should be lots of room to be improved in this algorithm. 

In this contribution, the basic concept of the timing-based localization and current performance and problems will be reviewed in the Sec.\ref{sec:localization}. In the Sec.\ref{sec:machinelearning}, a new approach for the timing-based localization utilizing the machine-learning technique will be demonstrated.

\section{LOCALIZATION BY THE CROSS-CORRELATION ANALYSIS}
\label{sec:localization}

As described in Ohno et al. 2018, the simulation framework for the timing-based localization for short GRBs has been developed, taking into account many observational conditions, such as the satellite altitude, detector response, shape of the observed GRB lightcurves. This simulation includes the satellite orbit simulation, full Monte-Carlo simulation of the detector, and the actually observed GRB light curves which is taken from the Fermi-GBM 3rd GRB catalog\cite{0067-0049-223-2-28}.
The time-delay for each satellite is calculated by the given GRB position and orbital parameters. The expected light curves by the CAMELOT detectors with given time-delay are simulated based on the Fermi-GBM light curves, scaling the detector response of the CAMELOT detector, including the Poissonian fluctuations. The time-delay of these simulated light curves are estimated by the cross-correlation function (CCF) analysis. A large number of simulation sets gives the mean value and its standard deviation of the time-delay for each satellite combination. Finally, the localization is performed by input the calculated time-delay and uncertainty into the following simple $\chi^2$ minimization formula,
\begin{equation}
\label{eq:chi2}
\chi^2 \equiv \sum_{i=0}^N \frac{\biggl\{\delta t_{\rm sim,i} - {\rm Norm} \times \rm{cos}\theta_{\rm model,i}  (R.A., Dec.) \times \rm D/c \biggr\}^2}{\sigma_{\rm sim,i}^2}, 
\end{equation}
where, $i$ is the number of the combinations of the visible satellites and $\sigma$ is the error of the simulated  $\delta t$.

The concept block diagram of this simulation framework and example of the simulated light curves can be found in the previous paper Ohno et al. 2018.
In Ohno et al. 2018, the localization uncertainty of this framework has been reported as 20 arcminuits for 13 \% of bright Fermi-GBM short GRBs. In this paper, more sophisticated and careful analysis and check have found that this estimation was a bit optimistic, especially in the procedure of estimation of the peak of the CCF. In the previous framework, the peak of the CCF has been estimated by the fitting by the gaussian function in an automated way, however it is found that this automation procedure was sometimes trapped into the local finer structure in the CCF, which does not come from the real time-delay of the two light curves. 

\begin{figure}[htbp]
\begin{center}
\begin{minipage}{8cm}
 \rotatebox{-0}{\resizebox{6cm}{!}{\includegraphics{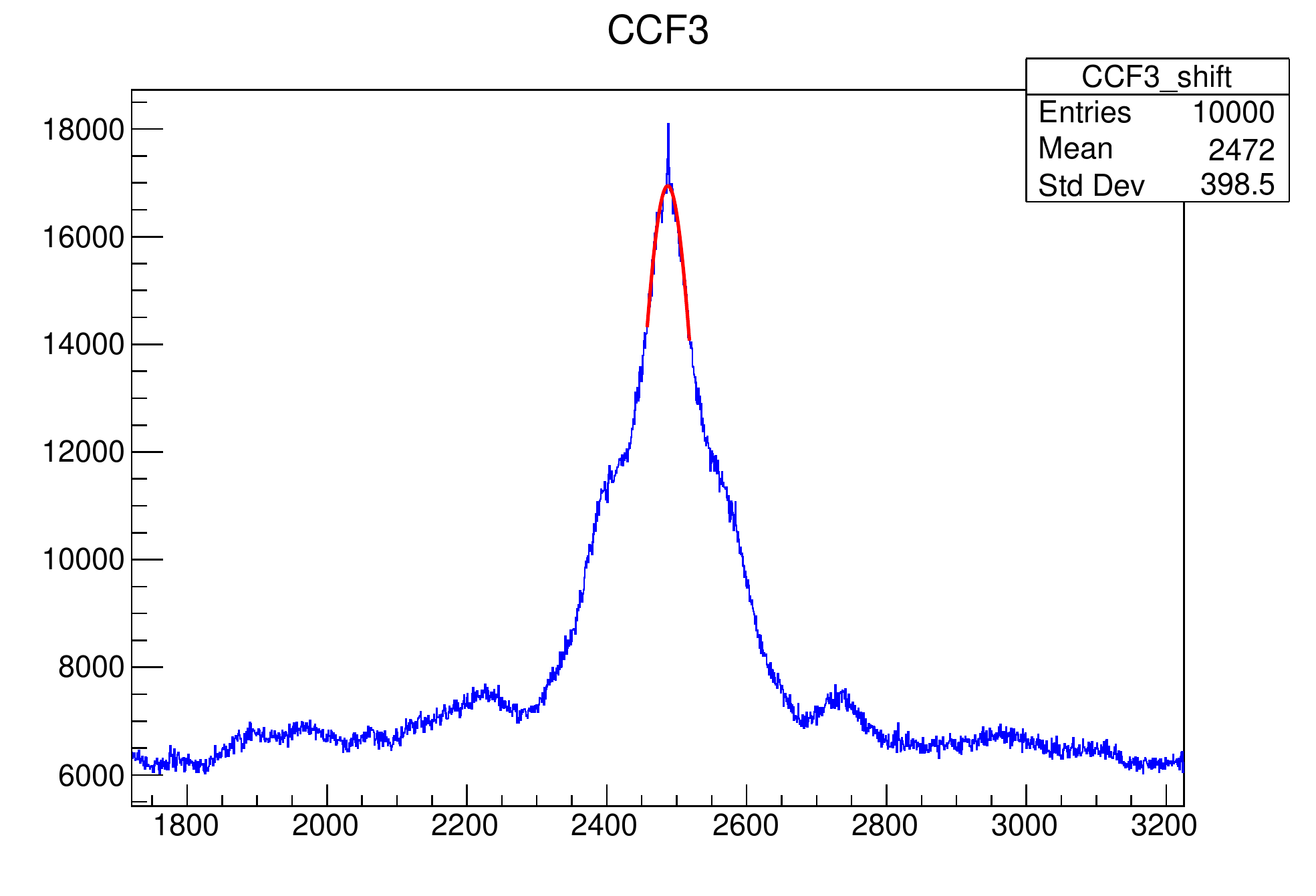}}}
\end{minipage}
\begin{minipage}{8cm}
 \rotatebox{-0}{\resizebox{6cm}{!}{\includegraphics{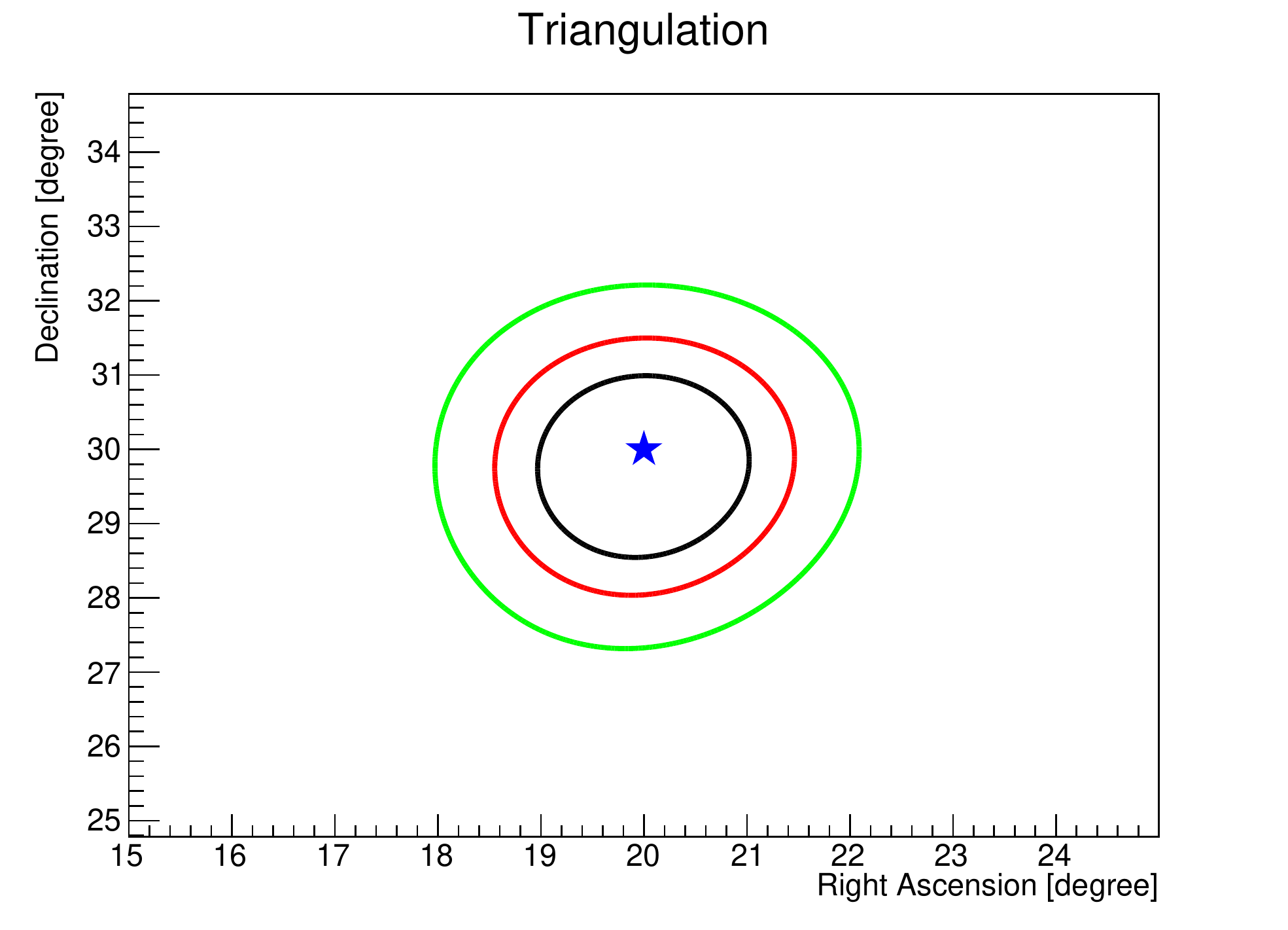}}}
\end{minipage}
\caption{(Left) The example of the cross-correlation function (CCF) calculated based on the bright short GRB 090227772. The red line shows the gaussian profile to fit around the peak of the CCF. (Right) The localization result of GRB 090227772 by using the parameterized $\chi^2$ fitting procedure assuming nine satellite constellations. The star marker is the assumed GRB position (R.A., Dec)=(20,30) degree and black, red and green contours represent the 1, 2, and 3-$\sigma$ confidence level, respectively}
\label{fig:CCFandLoc}
\end{center}
\end{figure}

This fitting procedure was carefully re-analyzed. Figure \ref{fig:CCFandLoc} left shows an example of the CCF calculated based on the bright short GRB 090227772, whose fluence was 1.1$\times$10$^{-5}$ erg cm$^{-2}$, and its localization uncertainty was reported as around 20 arcminutes in the previous framework. The global structure around the peak which should come from the real time-delay is well fitted by the gaussian profile. After such careful check of this fitting procedure and confirmed that there is no failed fitting to all of the CCFs, the parameterized $\chi^2$ minimization fitting in Eq.(\ref{eq:chi2}) for the obtained time-delay and its uncertainty has been performed again. Figure \ref{fig:CCFandLoc} right shows the result of the updated framework with a careful human-eyes inspection for the CCF analysis and the localization uncertainty got worse to be around 1 degree. This localization uncertainty is somehow consistent with that reported by other groups \cite{2019ExA....48...77W}. 

It should be addressed that even though the localization uncertainty was degraded from our previous paper, Ohno et al. 2018, the concept of this localization framework is still working well. The parameterized $\chi^2$ minimization method can localize GRBs once the time-delay and its uncertainty are estimated by any approaches.
The problem of this framework is that the CCF sometimes shows quite complicated structures which prevent us from developing the automated analysis framework. The automated analysis framework should be important for future large number satellite constellations.

\section{NEW APPROACH: MACHINE-LEARNING TECHNIQUE}
\label{sec:machinelearning}
As described in the previous section, our developed timing-based localization simulation framework based on the CCF analysis works well to localize the GRBs, but current framework is difficult and risky to operate in a fully automated way, mainly because of the complicated procedures for estimation of the peak of the CCF. In this section, a new approach utilizing the machine-learning technique is proposed and the very preliminary study and its performance are demonstrated. 
There are many parameters which can be estimated by the machine-learning technique, such as the peak of the CCF and the time-delay of two light curves.
The ultimate goal of this study would be to construct the neural network which returns the position of the GRB as the output layer with only light curves, the attitude and position of each satellite as the input. The simple time-delay estimation by the neural network would be a nice start of this study to confirm if the machine-learning approach is really helpful for the timing-based localization.
The basic idea of constructing neural network to predict the time-delay of two light curves and its application for the localization algorithm are followings;
\begin{enumerate}
\item Train the neural network by using a large number set of light curves with various shape patterns and various time-delay patterns.
\item Predict the time-delay of two simulated light curves based on the actually observed GRBs.
\item The mean and standard deviation of the predicted time-delay for a certain number of simulations are taken as the time-delay and its uncertainty
\item The obtained time-delay and uncertainty for each satellite combination is used for the parameterized $\chi^2$ minimization algorithm.
\end{enumerate}

\begin{figure}[htbp]
\begin{center}
 \rotatebox{-0}{\resizebox{16cm}{!}{\includegraphics{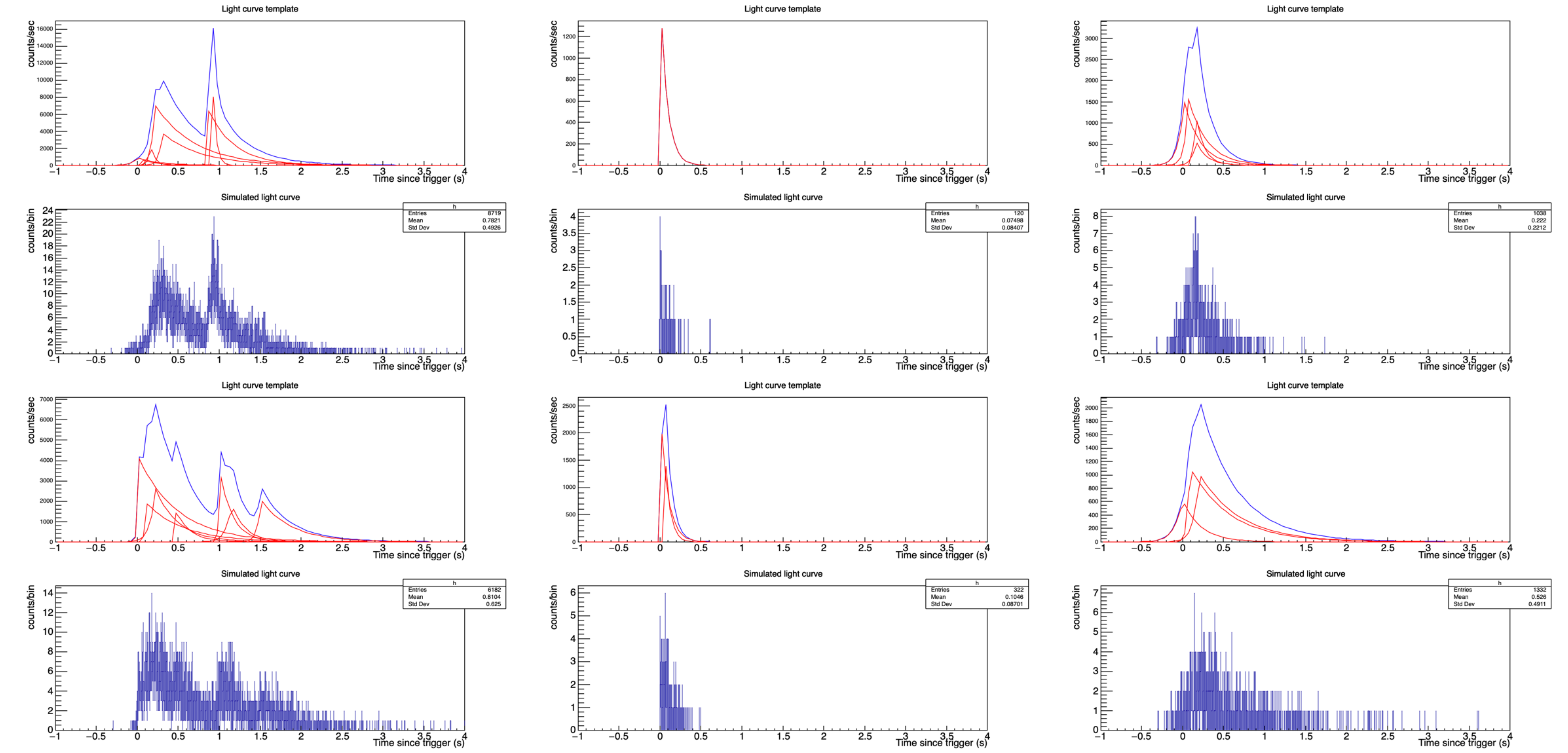}}}
 \caption{Examples of randomly generated light curve templates. The top part of each panel shows the template functions; each pulse is shown by red lines and the sum of those pulses shown by blue line. The vertical axis is in the unit of count per seconds. The bottom part of each panel shows example of simulated light curve in the 1-ms time bin based on the given template function. The vertical axis is in the unit of count per 1 ms. The horizontal axis of all panels shows the time in seconds.}
 \label{fig:lc_template}
 \end{center}
 \end{figure}

\begin{figure}[htbp]
\begin{center}
\begin{minipage}{8cm}
 \rotatebox{-90}{\resizebox{6cm}{!}{\includegraphics{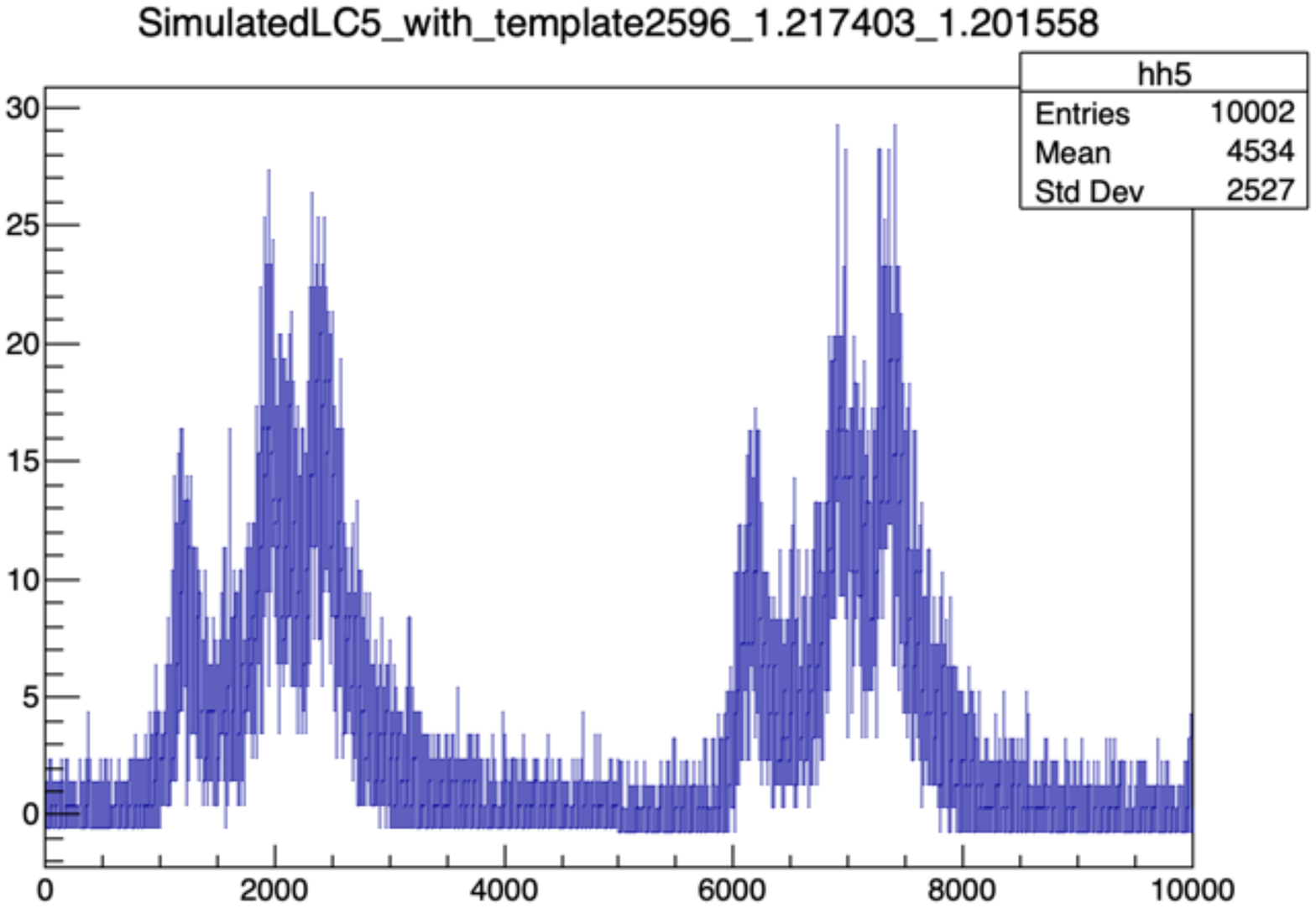}}}
\end{minipage}
\begin{minipage}{8cm}
 \rotatebox{-0}{\resizebox{7cm}{!}{\includegraphics{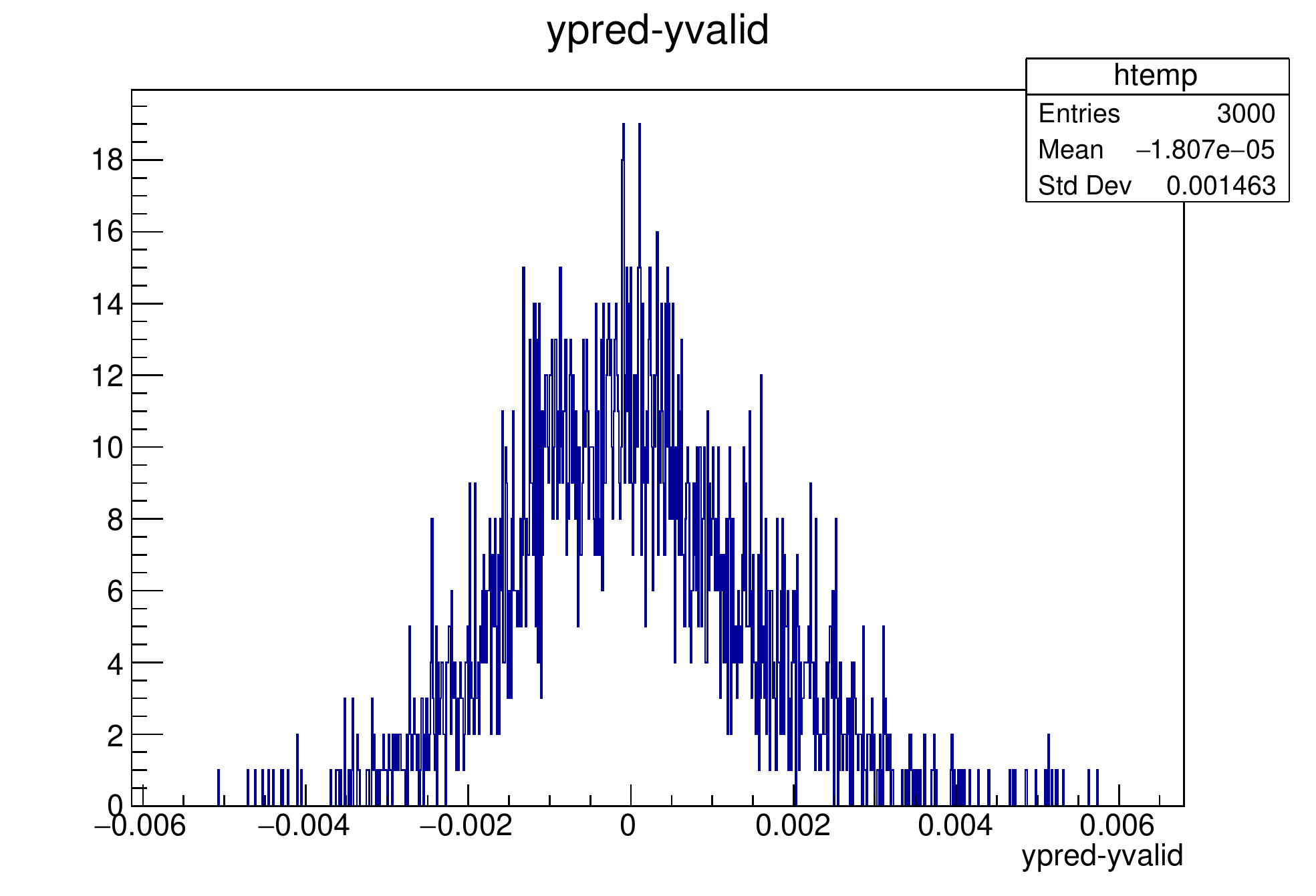}}}
\end{minipage}
\caption{(Left) Example of the light curve data for training of the neural network. Two light curves are combined in the 1-dimensinal shape. (Right) Distribution of the differences between input and predicted time-delay for a single light curve template by using the constructed neural network.}
\label{fig:lcsample}
\end{center}
\end{figure}

In the first step, a larger number of light curve patterns would be better in order to construct the well-trained network. However, the number of actually observed light curves is too small to use as the seed light curve template. Therefore, in this study the light curve template with various shape patterns have been generated by using the combination of the two exponential functions.
The single pulse can be drawn by,
\begin{eqnarray*}
    F(t) &=& A \times {\rm exp}(-(t_{\rm peak}-t)/\tau_1) ~~~~~~~{\rm for~~~ t<t_{\rm peak} }\\
         &=& A \times {\rm exp}(-(t-t_{\rm peak})/\tau_2) ~~~~~~~{\rm for~~~ t>=t_{\rm peak} }
\end{eqnarray*}
, where A, t$_{peak}$, $\tau_1$, and $\tau_2$ are the normalization, peak time, rise time constant and decay time constant of each pulse respectively, which are randomly generated. The final form of this function is the sum of the random number of those pulses. Figure \ref{fig:lc_template} shows examples of the generated light curve "template" together with the example of the simulated light curve based on this template. Various types of light curve shapes, such as a single fast rise and exponential decay pulse, separated pulses, and multiple complicated pulses are well generated by this method. 
In this study, 10,000 patterns of light curve shapes have been prepared and the
total 10$^6$ training data for the neural network have been generated based on 
randomly selected light curve template.
Each training data consists of a 1-dimensional linear combination of two light curves. One light curve does not contain the time-delay, but another contains a given time-delay. The time bin of the light curve is fixed to be 1 ms and covered from -1 s to +4 s from the trigger, which means the input layer of the neural network has 10$^4$ data points. The time-delay is also given randomly from $-$50 ms to +50 ms, which is the expected maximum time-delay considering our orbit. Figure \ref{fig:lcsample} left shows the example of the training light curve data.
The validation data is also generated by another set of light curve templates. Our neural network is constructed by the Keras tensorflow backend in the python library \cite{gulli2017deep,chollet2015keras}. The number of hidden layers set to be five and each hidden layer are fit by the sigmoid activation function and Adam optimizer as the first trial. Figure \ref{fig:lcsample} right shows an example of the performance of this trained neural network as the distribution of the differences between input and predicted time-delay for a single light curve template. As seen in the figure, the predicted time-delay shows a good agreement with the input value with a standard deviation of around 1.4 ms.

\begin{figure}[!h]
\begin{center}
 \rotatebox{-0}{\resizebox{17cm}{!}{\includegraphics{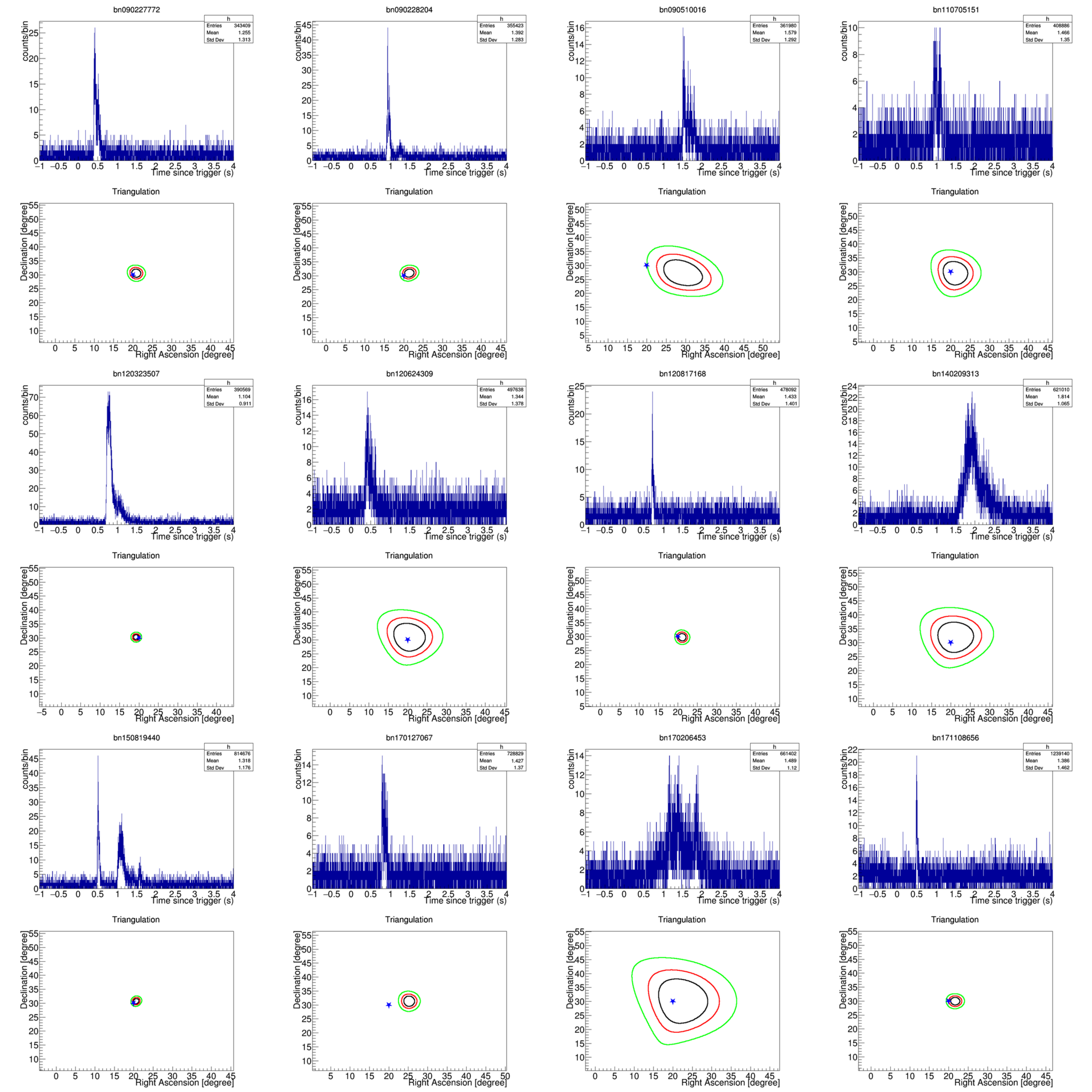}}}
 \caption{Localization result of 12 bright short GRBs by using the machine-learning approach to estimate the time-delay of each satellite combination. Top and bottom part of each panel shows the light curve of each GRBs in 1 ms time scale, and the localization contour same as the fig \ref{fig:CCFandLoc} right.}
 \label{fig:loc_ML}
 \end{center}
 \end{figure}

This trained neural network is used to predict the time-delay of the real GRB data. The light curve is simulated using the real GRB data as the template, which is the same procedure as described in Sec.\ref{sec:localization}. The mean value and standard deviation are obtained from 3000 realizations for each satellite combination. Now, the time-delay and its uncertainty are ready to input the same parameterized $\chi^2$ minimization algorithm in the Sec.\ref{sec:localization}. Figure \ref{fig:loc_ML} shows the result of the localization for 12 bright short GRBs observed by the Fermi-GBM. First of all, this machine-learning approach successfully works to estimate the time-delay of each satellite combination, and the predicted time-delay well localizes the GRBs with 1-degree uncertainty for bright short GRBs and $\sim$5-degree uncertainty for intermediate brightness short GRBs. The training procedures are completely performed by the artificially generated light curve data, which means that this trained neural network can be applied for any kind of newly detected bright GRBs. Furthermore, once we construct the trained neural network, the procedures to estimate the time-delay of two light curves can be fully automated. Therefore, this machine-learning approach for the timing-based localization would be very promising. Further detailed study of this machine-learning approach and application for the timing based localization will be reported in the separated paper.

\clearpage

 \clearpage
\acknowledgments 
 The research has been supported by the European Union, co-financed by the European Social Fund (Research and development activities at the E\"{o}tv\"{o}s Lor\'{a}nd University's Campus in Szombathely, EFOP-3.6.1-16-2016-00023).

\bibliography{mybibfile_v20180622.bib} 
\bibliographystyle{spiebib} 

\end{document}